# The composition of M-type asteroids II: Synthesis of spectroscopic and radar observations

J.R. Neeley,[a,b,1] B.E. Clark,[a,1] M.E. Ockert-Bell,[a,1] M.K. Shepard,[c,2] J. Conklin,[d] E.A. Cloutis,[e] S. Fornasier,[f] S.J. Bus[g]

[a] Department of Physics, Ithaca College, Ithaca, NY 14850
[b] Department of Physics, Iowa State University, Ames, IA 50011
[c] Department of Geography and Geosciences, Bloomsburg University, Bloomsburg, PA 17815
[d] Department of Mathematics, Ithaca College, Ithaca, NY 14850
[e] Department of Geography, University of Winnipeg, Winnipeg, MB, R3B 2E9
[f] LESIA, Observatoire de Paris, 5 Place Jules Janssen, F-92195 Meudon Principal Cedex, France
[g] Institute for Astronomy, 2680 Woodlawn Dr., Honolulu, HI 96822

[1] Guest observer at the NASA Infrared Telescope Facility
[2] Guest observer at Arecibo Observatory

Email: jrneeley@iastate.edu

**Abstract**

This work updates and expands on results of our long-term radar-driven observational campaign of main-belt asteroids (MBAs) focused on Bus-DeMeo Xc- and Xk-type objects (Tholen X and M class asteroids) using the Arecibo radar and NASA Infrared Telescope Facilities (Ockert-Bell et al. 2008; 2010; Shepard et al. 2008; 2010). Eighteen of our targets were near-simultaneously observed with radar and those observations are described in Shepard et al. (2010). We combine our near-infrared data with available visible wavelength data for a more complete compositional analysis of our targets. Compositional evidence is derived from our target asteroid spectra using two different methods, a $\chi^2$ search for spectral matches in the RELAB database and parametric comparisons with meteorites. We present four new methods of parametric comparison, including discriminant analysis. Discriminant analysis identifies meteorite type with 85% accuracy. This paper synthesizes the results of these two analog search algorithms and reconciles those results with analogs suggested from radar data (Shepard et al. 2010). We have observed 29 asteroids, 18 in conjunction with radar observations. For eighteen out of twenty-nine objects observed (62%) our compositional predictions are consistent over two or more methods applied. We find that for our Xc and Xk targets the best fit is an iron meteorite for 34% of the samples. Enstatite Chondrites were best fits for 6 of our targets (21%). Stony-iron meteorites were best fits for 2 of our targets (7%). A discriminant analysis suggests that asteroids with no absorption band can be compared to iron meteorites and asteroids with both a 0.9 and 1.9 µm absorption band can be compared to stony-iron meteorites.

## 1. Introduction

Main-belt asteroids (MBAs) are critical for testing and modifying formation models of the Solar System. X-complex asteroids (Tholen, 1984; Bus and Binzel, 2002b) play a fundamental role in the investigations of early formation theories because they represent about 20% of inner main-belt asteroids (Mothe-Diniz et al., 2003) and because they potentially include metallic asteroid cores (cf. Bell et al., 1989).

We started this work in 2003, targeting the Tholen M-types (Tholen, 1984; Tholen and Barucci, 1989). Since that time, taxonomy has evolved and our targets have been newly classified according to the Bus-DeMeo system (DeMeo et al., 2009). The Bus-DeMeo taxonomy builds on the Bus taxonomy (Bus, 1999; Bus and Binzel, 2002a,b; DeMeo et al., 2009) by using both visible and near-infrared data to group asteroids spectrally from 0.4 to 2.5 μm according to principal components. In the Bus-DeMeo system, our targets are now primarily called Xc- and Xk-types, however some of them are designated Xe or X-types. These asteroids have a notable lack of strong spectral features, and their near-infrared slopes range widely from flat to red. X-class asteroids have high slope values between 0.2 and 0.38 reflectance/μm. Xk-types are characterized as having a slight absorption near 0.9 μm. Xc-types have a low to medium slope and are slightly curved and concave downward (DeMeo et al. 2009).

Because they have very few and very subtle spectral features, our survey of the X complex objects combines radar measurements with visual and near-infrared wavelength measurements for a more complete compositional determination. First results were presented in Ockert-Bell et al. (2008) and Shepard et al. (2008). In Ockert-Bell et al. (2010) we showed that six of seven asteroids with the highest iron abundances had consistent compositional predictions between our spectral techniques and radar evidence. For three of seven asteroids with low metal abundance, our spectral results were consistent with the radar evidence. The remaining seven asteroids had ambiguous compositional interpretations when comparing the spectral analogs to radar analogs.

In this paper, we present updated results from our long-term survey of the X-complex asteroids. We present new observations (69 Hesperia, 325 Heidelberga, 336 Lacadiera, 337 Devosa, 516 Amherstia, 792 Metcalfia, and 1214 Richilde), new comparisons of asteroid and meteorite spectra, parametric comparisons of asteroid and meteorite spectra, a new discriminant analysis, and a synthesis of all the work we have conducted to date. This paper directly follows and builds on our previous results (Shepard et al., 2010; Ockert-Bell et al., 2010) and corrects several typographical errors found in Ockert-Bell et al. (2010) post-publication (see Appendix A).

## 2. Observations and data reduction

Our observations were conducted at the Mauna Kea Observatory 3.0 m NASA Infrared Telescope Facility (IRTF) in Hawaii. We used the SpeX instrument, equipped with a cooled grating and an InSb array (1024 x 1024) spectrograph at wavelengths from 0.82 to 2.49 μm (Rayner et al., 2003). Spectra were recorded with a slit oriented in the north-south direction and opened to 0.8 arcsec. A dichroic turret that reduces the signal below 0.8 μm was used for all observations.

**Table 1** gives a summary of the asteroids that were observed in the near-infrared and radar for this program and **Table 2** gives the observing circumstances. Following normal data reduction procedures of flat fielding, sky subtraction, spectrum extraction, and wavelength calibration, each spectrum was fitted with the ATRAN atmospheric model for telluric absorption features (Lord, 1992; Bus et al., 2003; Sunshine et al., 2004). This procedure required an initial estimate of precipitable water in the atmospheric optical path using the zenith angle for the observation and the known τ-values (average atmospheric water) for Mauna Kea. This initial guess was iterated until the best fit between predicted and observed telluric band shapes was obtained, and an atmospheric model spectrum was generated (Bus et al., 2003). Following this, each asteroid spectrum was divided by the atmospheric model and then ratioed to each star spectrum, similarly reduced, before normalization at 1.2 μm. The final spectra we report are averages of 3-5 asteroid/star ratios, calculated to average out variations due to standard star and sky variability.

Usually, 2-5 different standard stars were observed on any given night at the telescope (64 Hyades, or any of 13 Landolt stars). We used only solar standard stars. In addition, 1-3 observations were obtained of each different standard star. Although we did not attempt a quality judgment of the asteroid observations, we paid careful attention to the star observations in order to eliminate "bad" spectra and/or "bad" nights at the telescope (such as nights when unresolved stratospheric clouds were too variable to be removed in reduction). As a matter of routine, we calculate the ratios of asteroid1/star1 over asteroid1/star2 for all permutations of 4-5 stars per night to help find bad star measurements (see discussion in Ockert-Bell et al., 2008).

## 3. Data analysis

**Figs. 1-4** present our average asteroid spectra over all aspects. When possible, we include visible data obtained by other workers (Bus and Binzel, 2002a; Zellner et al., 1985; Chapman and Gaffey, 1979; Fornasier et al., 2010). Error bars are not plotted; however formal measurement uncertainties that propagate through the reduction are less than the scatter in the data, so we suggest that the scatter in the data is the best illustration of the uncertainties of each spectrum measurement. **Table 3a** presents a summary of the spectral characteristics of our asteroids and radar results (whether the radar albedo indicates high, moderate or low NiFe content). Albedos are taken from the JPL Horizons database or Shepard et al. (2010).

*3.1 Spectral features*
In this paper, we will refer to several spectral features seen in our dataset. In the near-infrared we see the 0.9 µm (Band 1) and 1.9 µm (Band 2) absorptions attributed to mafic minerals, in particular ortho- and clinopyroxenes (Adams, 1974; Hardersen et al., 2005). The visible (VIS) region from 0.45 to 0.7 µm was chosen to represent the visible continuum slope before a peak brightness at about 0.75 µm. The near-infrared slope regions were selected because there is a change in slope at about 1.5 µm. The precise wavelength of slope change varies from 1.35 to 1.65 µm and so the two NIR regions are separated by 0.15 µm. Near-infrared slope (NIR1) is from 1.1 to 1.45 µm, and NIR2 is from 1.6 to 2.3 µm. The slope values are robust; changing the boundaries by 0.02 µm changes slope values by less than 2%.

Detection of the mafic bands is difficult due to the fact that they are co-located in wavelength with atmospheric water bands. If our Band 1 and 2 detections were due to atmospheric water, then we might also expect to detect an absorption at 1.4 µm, which is also strong. The fact that we do not detect an absorption at 1.4 µm, suggests that our atmospheric corrections are appropriate. Hardersen et al., 2005 show that the shape of M-type asteroid bands are quite different from those of atmospheric water and argue that they are, instead, weak mafic bands.

When an absorption feature is detected, band centers and band depths are calculated. First a linear continuum is fitted at the edges of the band; that is from the peak in the visible region to the region between 1.3 and 1.4 µm. Then, the spectra are divided by the linear continuum and the region of the band is fitted with a polynomial of order between 2 and 7. The band center is calculated as the position where the minimum of the polynomial occurs, and the band depth as the minimum in the spectral reflectance curve relative to the fitted continuum.

To indicate the strength of the band, we calculate a quantity that we call the "Band Signal Strength Ratio": we divide the depth of the band by the standard deviation between the polynomial fit and the reflectance data. In Ockert-Bell et al. (2010) we carefully compared band signal strength ratios for two cases for each asteroid: when the band was fit with a straight line versus when the band was fit with a polynomial curve. Band signal strength ratio values measured from 1.5 to 10 and were compared with meteorite bands measured similarly. This process helped us to establish a threshold value of approximately 3 for a secure band detection (i.e. when a band is present at the 3σ level). If a band signal strength ratio exceeds 3, then we feel confident that we have a detection. For values below 3, we suggest the data are either too noisy, or the band is too faint for detection. The band depths and band signal strength ratios for each asteroid are given in **Table 3b**. We note that threshold values for secure detection depend on many factors and should be individually determined for each data set. In Clark et al. (2011) the data supported a secure detection at threshold values of 10 (i.e. when a band is present at the 10σ level).

The spectral feature measurements are presented in **Table 3a**. **Table 3a** sorts our targets according to the presence of the mafic bands, creating four groups: Group 1 (**Fig. 1**): no absorption bands, Group 2 (**Fig. 2**): 0.9 μm band only, Group 3 (**Fig. 3**): 0.9 and 1.9 μm bands detected. One object, 785 Zwetana, does not fall into any of the above groups (**Fig. 4**) since the location of the detected bands is significantly different from all other targets.

*3.1.1. Group 1: no mafic bands*
**Fig. 1** presents asteroid target spectra that do not appear to show either the mafic band at 0.9 μm or at 1.9 μm. The only distinguishing characteristic in the near-infrared for these spectra is the position of the change in slope in the near-infrared at approximately 1.5 μm. In addition to 21 Lutetia and 97 Klotho as presented in Ockert-Bell et al. (2010), we have added 325 Heidelberga to the group 1 asteroids.

*3.1.2 Group 2: measurable 0.9 μm but not 1.9 μm*
**Fig. 2** presents the asteroid spectra that have measurable absorption bands at 0.9 μm but not at 1.9 μm. The band centers for these objects range from 0.80 to 0.96 μm. Ockert-Bell et al. (2010) placed 16 Psyche, 22 Kalliope, 77 Frigga, 129 Antigone, 135 Hertha, 136 Austria, 250 Bettina, 441 Bathilde, 497 Iva, 678 Fredegundis, 771 Libera, and 872 Holda in group 2. However, applying our new band detection threshold, we find that 224 Oceana, 758 Mancunia, and 779 Nina should be reclassified into group 2. From new observations we add 69 Hesperia, 336 Lacadiera, 337 Devosa, 792 Metcalfia, and 1214 Richilde to group 2.

*3.1.3 Group 3: measurable 0.9 μm and 1.9 μm*
**Fig. 3** presents the asteroid spectra with measurable 0.9 and 1.9 μm mafic absorptions. These objects have the characteristic spectrum of an Xk asteroid with a strong red continuum slope. The band centers of the 0.9 μm band for these objects range from 0.89 to 0.93 μm. The band centers of the 1.9 μm band range from 1.72 to 1.95 μm. From Ockert-Bell et al. (2010) we have 55 Pandora, 110 Lydia, 216 Kleopatra, and 347 Pariana. From new observations we include 516 Amherstia.

*3.1.4 The oddball: different from everything else*
Zwetana does not have measurable mafic bands and has a concave curvature in the near-infrared (**Fig. 4**). Of the objects in our SpeX database, few have the strong red slope with concave curvature similar to Zwetana's spectrum. Two X objects (229 Adelinda and 276 Adeleid (Clark et al., 2010)) and a T object (517 Edith) have similar concave shape but there are no SMASS data for any of these objects. See Ockert-Bell et al. (2010) for a more complete discussion of Zwetana.

**4. Discussion**

*4.1 Search of RELAB spectral database for meteorite analogs*
To constrain the possible mineralogies of our targets, we conducted a search for meteorite and/or mineral spectral matches. We used the publicly available RELAB spectrum library (Pieters, 1983), which consisted of nearly 15,000 spectra in the November of 2008 public download. We added spectra for iron meteorites presented in Cloutis et al. (2010), mixtures of iron meteorites and pyroxenes (Cloutis et al., 2009), carbonaceous chondrites, and Vaca Muerta (Ed Cloutis, personal communication). For each spectrum in the library, a filter was applied to find relevant wavelengths (0.4-2.5 μm). A second filter was applied to reject spectra with brightness at 0.55 μm that differed from the asteroid's albedo by more than ±7% (absolute). For example, if the asteroid's albedo was 10%, then we compared it to all meteorite spectra with brightnesses from 3% to 17%. This process produced a list of approximately 4000 RELAB spectra of appropriate brightness and wavelength coverage for comparison to the asteroid.

The "albedo" (brightness) of a meteorite is taken to be the reflectance value at 0.55 μm of a sample measured at 30° phase angle. This value is compared with the asteroid geometric albedo ($p_v$) at a wavelength of ~0.55 μm (usually V band at 0° phase angle). While this comparison is decidedly inexact, we have never seen an empirical "correction" factor that can be used to convert reliably from meteorite reflectance to asteroid albedo. It is certain that the two measurements are correlated, and that they can be compared in a relative sense. Meteorite reflectance varies with grain size and sample packing (an effect we have tried to account for by including various grain sizes in our comparison meteorite dataset), and asteroid albedos are generally known to within 20% in absolute terms (Tedesco et al. 2002). Considering these facts, we have decided to use published asteroid albedos and published meteorite reflectances without making any corrections from one quantity to the other, in order to make the least number of assumptions about the nature of the correlation between these quantities. Our large range in meteorite brightness of ±7%, is however, an attempt to allow for the inexact nature of the comparison of asteroid albedo & meteorite brightness.

We normalized our asteroid and meteorite spectra to 1.0 at 0.55 μm before calculation of the $\chi^2$ value for each RELAB sample relative to the input asteroid spectrum. The lists of RELAB spectra were sorted according to $\chi^2$, and then visually examined for dynamic weighting of spectral features by the spectroscopist. Given similar $\chi^2$ values, a match that mimicked spectral features was preferred over a match that averaged through small spectral band features. We visually examined the top ~50 $\chi^2$ matches for each asteroid to make sure that the results of our comparison algorithm were reasonable.

The search algorithm ranged over the entire database without constraints on particle size. Many of the matches found by our search algorithm are particulate materials, which are relevant for asteroid regoliths. However, the RELAB database does not contain very many spectra of particulate metal samples, and the search algorithm often resulted in solid-surface samples, such as slabs. While these surfaces are not considered to be relevant for asteroid regoliths, it is interesting that their spectral properties are a better match for many of the asteroids we observed. The two meteorites with the most matches, Landes and Esquel, are both slab samples. Cloutis et al. (2010) examine the systematic differences between particulate and slab samples of iron meteorites. Because it is rich in contextual information, we provide here a direct quote from their abstract:

> "The 0.35-2.5 μm reflectance spectra of iron meteorite powders and slabs have been studied as a function of composition, surface texture (for slabs), grain size (for powders), and viewing geometry (for powders). Powder spectra are invariably red-sloped over this wavelength interval and have a narrow range of visible albedos (~10-15% at 0.56 μm). Metal (Fe:Ni) compositional variations have no systematic effect on the powder spectra, increasing grain size results in more red-sloped spectra, and changes in viewing geometry have variable effects on overall reflectance and spectral slope. Roughened metal slab spectra have a wider, and higher, range of visible albedos than powders (22-74% at 0.56 μm), and are also red-sloped. Smoother slabs exhibit greater differences from iron meteorite powder spectra, exhibiting wider variations in overall reflectance, spectral slopes, and spectral shapes. No unique spectral parameters exist that allow for powder and slab spectra to be fully separated in all cases. Spectral differences between slabs and powders can be used to constrain possible surface properties, and causes of rotational spectral variations, of M-asteroids." (Cloutis et al. 2010)

**Table 4** shows the results of the search using normalization at 0.55 μm. **Fig. 5** shows the plots of the asteroid spectra with the spectra of their best matches plotted in red. Several asteroids were difficult to match with RELAB spectra partly because their albedos are greater than 0.20 (55 Pandora, 129 Antigone, 250 Bettina, and 678 Fredegundis). For these asteroids, it was difficult to find anything in the RELAB database with corresponding brightness. The closest matches in albedo and spectral shape were lunar samples. However, lunar samples have deeper 0.9 and 1.9 μm absorptions, which were inconsistent with the asteroid spectra.

Three meteorites showed up as best matches for many of the asteroids when the data were normalized to 0.55 μm. These meteorites are discussed below; they all have a substantial number of silicate inclusions and do not have the nearly pure metal composition of iron meteorites like Odessa.

The iron meteorite Landes was a spectral match for eleven asteroids, and the best match for seven. Landes is an iron meteorite with silicate inclusions and a large number of minerals in a NiFe matrix. Chemical analysis shows this meteorite to be 81% NiFe and 16% silicates (Bunch et al., 1972). This result may indicate that Landes-type material is more common on M-type asteroid surfaces than is pure NiFe metal.

The stony-iron meteorite Esquel was a spectral match for eight asteroids, and the best match for three asteroids. Esquel is described as containing "Large olivine masses transected by thin veins of metallic Fe-Ni (Ulff-Moller et al., 1998). The Ulff-Moller chemical analysis found 63% olivine and 21% NiFe.

The EH5 enstatite chondrite St. Mark's was a match for five and the best match for three asteroids. Enstatite chondrites have oxygen isotope compositions that plot near the terrestrial fractionation line, and highly reduced mineral assemblages (containing little FeO, Si-bearing metal, and sulfides of elements normally considered lithophile). The high-iron (EH) chemical group is distinguished by small chondrules (0.2 mm), abundant metal (~10 vol.%) that is rich in Si (~3 wt.%), and an extremely reduced mineral assemblage including niningerite (MgS) and perryite (Fe-Ni silicide). The Type 5 designation means that St. Mark's has been metamorphosed under conditions sufficient to homogenize olivine and pyroxene, convert all low-Ca pyroxene to orthopyroxene, cause the growth of various secondary minerals, and blur chondrule outlines (The Meteoritical Society, 2010).

*4.2 Parameter Comparisons*

The values of the six parameters (brightness, VIS slope, NIR1 slope, NIR2 slope, Band 1 center, and Band 2 center) of the asteroids can be compared with those of each of the meteorite classes (**Table 5**). Meteorite parameter values can be found in Tables 6a-d of Ockert-Bell et al. (2010). We use four tests and a discriminant analysis method to determine the meteorite class that is most compatible with each of our targets.

*4.2.1 Parameter Range Comparison Test*

We find the minimum and maximum range of each parameter for each meteorite group (e.g. the range in brightness for enstatite chondrites (EC) is 0.06-0.18). Asteroids were tested one-by-one in a binary sense (0 for not within meteorite parameter range, 1 for within range). We then added up the number of positive results (the number of 1s). The meteorite type that matched the greatest number of asteroid parameters was called the best fit type. Since many of the meteorite parameter value ranges significantly overlap, it is difficult to determine compositional predictions using only this method. The results of the comparison test are given for each asteroid in **Table 6**.

*4.2.2 Average Distance Test*

We find the absolute value of the difference between an asteroid's parameter value and the average parameter value of meteorites within each meteorite class. For each parameter, the best-fit type is determined by the lowest distance from the class average value. The class that appears most often between the six parameters is the overall best fit type. The results of the average distance calculation test are given for each asteroid in **Table 7**.

*4.2.3 Mean Distance Test*

We find the mean parameter value for each meteorite type. The absolute value of the difference between an asteroid's parameter value and the mean value for each meteorite type is calculated and divided by the standard deviation of the meteorite parameter values. We then sum the mean distances over all six parameters. The meteorite type with the lowest total sum was classified as the best fit type. The results of the mean distance test are given for each asteroid in **Table 7**.

*4.2.4 Median Distance Test*

For the parameters Band 1 center and Band 2 center, the mean parameter is unrealistic, since spectra with no band are given a value of zero. In these cases, using the median parameter value more accurately describes the actual parameter values of the meteorites. Therefore, in this test we find the median parameter value for each meteorite type. The absolute value of the difference between an asteroid's parameter value and the median value for each type is calculated. We then sum the median distances over all six parameters. The meteorite type with the lowest sum was selected as the best fit type. The results of the median distance test are given for each asteroid in **Table 7**.

*4.2.5 Discriminant Analysis*

Discriminant analysis is a statistical technique concerned with the relationship between a categorical variable and a set of interrelated variables. This technique has been widely used as a classification system for terrestrial geological studies (Prelat, 1977; Clausen and Harpoth, 1983). In this situation the categorical variable denotes membership in 8 classes of meteorites that we have designated (CICM, CKCOCV, CR, EC, IM, SIM and UR) and our variables were the 6 measurements of albedo, visual slope, NIR1 slope, NIR2 slope, Band 1 center and Band 2 center for each spectrum. The idea of R.A. Fisher's approach to discriminant analysis (MacLachlin, 2004) is to find the linear combinations of the measurements (the discriminant functions) that maximize the ratio of between-group to within-group sum of the squares (the analysis of variance F statistic). These discriminant (linear) functions can then be used to predict group membership of new objects (our asteroids) based on the values of the same independent variables. One limitation of the method is that this notion of separation between groups is based on an assumption that the values of the measurements are normally distributed (MacLachlin, 2004). In our case, Band 1 center and Band 2 center are not normally distributed since meteorites without an absorption band are given values of zero.

For training data to determine the discriminant functions we used the data from our collection of RELAB meteorite spectra. Using the statistical software package MINITAB, discriminant analysis was run using albedo, visual slope, NIR1 slope, NIR2 slope, Band 1 center and Band 2 center as the six independent variables. The analysis was run multiple times with different meteorite classes grouped together, in order to determine which combinations resulted in the greatest proportion of correct classification of the meteorites. We determined that it is best to group the CI and CM classes together, and the CK, CO, and CV classes together, instead of keeping all meteorite classes separate. This resulted in a total of eight groups (CICM, CKCOCV, CR, EC, IM, SIM, and UR). Out of 132 CI & CM meteorites, 121 were correctly identified as belonging to the CICM group. Out of 85 CK, CO, and CV meteorites, 72 were correctly identified. Out of 15 CR meteorites, 8 were correctly identified. Out of 9 EC meteorites, 6 were correctly identified. Out of 23 IM meteorites, 17 were correctly identified. Out of 5 SIM meteorites, 4 were correctly identified. Out of 19 UR meteorites, 18 were correctly identified. Overall, the discriminant functions correctly classified the meteorites 85.4% of the time. This is summarized in **Table 8a**. The discriminant functions did not perform as well for two classes (CR and EC). This is likely due to low number statistics. The known meteorites within these classes have too much variation within our measurable parameters to be well classified.

The most likely meteorite type predictions for our asteroids are given in **Table 8b**. The probability that each prediction is reliable is also given. In cases where the probability was less than 50%, we list the two most likely predictions as a best-fit type.

*4.3 Sythesis of parametric analysis*
The results of parameterization, four comparison tests and a discriminant analysis are summarized in **Table 9**. In **Table 9** we show the best-fit types for each asteroid for each method, an overall best-fit type for the parametric tests, and the RELAB best-fit type. The overall best-fit type for parametric tests is given when a meteorite type is a best-fit for three or more parametric comparison methods.

**5. Synthesis with radar**

As discussed in Shepard et al. (2010), radar may be a better tool than spectroscopy for identifying metallic content in asteroids because the spectra of metallic minerals are mostly featureless. The assumption in the work of Shepard et al. (2010) is that iron is several times denser than rock-forming silicates and, since radar reflectivity depends primarily upon near-surface bulk density (Ostro et al., 1985), an asteroid composed primarily of iron will have a significantly higher radar reflectivity than one composed of silicates.

Shepard et al. (2010) developed a heuristic model for linking the observed radar albedo to an asteroid's near-surface bulk density in order to determine the best meteoritic analog(s) for M-type asteroids. To do this, they used what is known about the radar reflectivity of powdered materials and the consensus view of MBA surfaces and compositions. The model estimates surface bulk density from radar albedo (or vice versa), and allows comparison of the estimated surface densities to those of meteorite analogs (Britt and Consolmagno, 2003). For consistency with Shepard et al., we considered iron meteorites (IM), low and high metal enstatite chondrites (EC), high metal carbonaceous chondrites (in particular CH and CB), low metal carbonaceous chondrites (CI and CM), and stony-iron meteorites (SIM).

Based on the radar observations, the model NiFe content (high, moderate, low) for each of the asteroids in our survey is copied from Shepard et al. (2010) into **Table 3a**. In Ockert-Bell et al. (2010) we concluded that the highest metal-content asteroids tend to exhibit both absorption features at 0.9 and 1.9 μm, and the lowest metal-content asteroids tend to exhibit either no bands or only the 0.9 μm band. However, as seen in **Table**

**3a**, this conclusion no longer holds. Of the eight asteroids with the highest metal content as indicated by radar, five exhibit only the 0.9 μm band and the remaining three exhibit both the 0.9 and 1.9 μm bands.

The best radar analogs for each of the asteroids in our survey (Shepard et al. (2010, 2011)) are presented in **Table 10** and compared with the results from our RELAB searches, parametric comparison tests, and discriminant analysis.

We find that for four of the eight asteroids with the highest iron abundances, our spectral results are consistent with the radar evidence. For three of the eight with the lowest metal abundances, our spectral results are consistent with the radar evidence. The remaining twelve asteroids have ambiguous compositional interpretations when comparing the spectral analogs with the radar analogs. The number of objects with ambiguous results from this multi-wavelength survey indicates perhaps a third diagnostic wavelength region (such as the mid-infrared around 2-4 μm, the mid-infrared around 8-25 μm, and/or the ultraviolet around 0.2-0.4 μm) should be explored to resolve the discrepancies.

**6. Summary**
With the inclusion of seven additional asteroids to our near-infrared observation database, bringing our total to 29 asteroids, we are able to update and expand on some of the results from Ockert-Bell et al. (2010).
   (1) Discriminant analysis uses a linear combination of parameter values to correctly classify meteorite groups 85% of the time.
   (2) Twenty-six of twenty-nine asteroids in our survey have weak absorptions at 0.9 μm and six of twenty-nine have a weak absorption at 1.9 μm. By definition, these asteroids would be Xk asteroids in the Bus-DeMeo taxonomy.
   (3) Three of the eight asteroids with the highest modeled iron content from radar studies have both the 0.9 and 1.9 μm absorption features. These absorptions indicate the presence of pyroxene, so these asteroids are not consistent with pure nickel-iron metal surfaces.
   (4) Few of the iron and enstatite chondrite meteorite analogs exhibit the 1.9 μm absorption. For a near-infrared spectrum of pyroxene-rich material, increasing amounts of metal will cause both the 0.9 and 1.9 micron bands to weaken. However, the 1.9 μm band will disappear first as metal abundance increases. We find it surprising and odd that we detect a weak 1.9 μm feature in the asteroid spectra and yet find no evidence for the presence of the feature on many of the analog meteorites. There are many possible explanations and we list three of them here: (1) We did not completely remove telluric absorption features. We suggest that this is unlikely because the strong telluric feature at 1.4 μm is not present in our asteroid spectra. (2) Mesosiderites and CH meteorites are the best available analogs for the asteroid surface but are not represented well in the RELAB database and hence do not have a chance to show up in our RELAB searches or statistical methods. (3) The terrestrial meteorite collection is incomplete and we do not have analogs in our meteorite collections for Group 3 asteroid surface materials. In this case, a mineralogical study can help find possible analogs.
   (5) A search for meteorite spectral matches in the RELAB library found that twelve of twenty-nine asteroids could be modeled using the Landes iron meteorite. For seven of twenty-nine Landes was the best match. Asteroids in our sample with very high visible albedos were not well matched (except by lunar samples, which have deeper absorptions and are not appropriate analogs for cosmochemical reasons).
   (6) A parametric search of the asteroid and meteorite samples shows that the best matches for our asteroids are usually iron meteorites (**Table 9**). However, we note that the measurements of iron meteorites found in the RELAB database are most often slab measurements and that a number of the iron meteorites measured contain significant inclusion of pyroxenes and other minerals.
   (7) For twenty-five of twenty-nine asteroids (86%), our parametric comparison test results were consistent with each other in three out of four methods in determining a likely meteorite analog type.
   (8) For seven of eighteen asteroids (39%) one or more of our meteorite analog type predictions were consistent with radar evidence. Ockert-Bell et al. (2010) reported 40% agreement between radar evidence and one or more spectral method. Hence, our modest increase in sample size has not significantly changed the rate of agreement between methods of compositional deconvolution.

(9) Discriminant analysis indicates that asteroids with no silicate bands are comparable with iron meteorites and that asteroids with both a 0.9 and 1.9 µm absorption band can be compositionally compared to stony-iron meteorites (**Table 8b**).
(10) This paper synthesizes results from three methods: RELAB spectral library searches, parametric comparison tests with meteorites, and radar albedo bulk density analogs. Our curve-matching spectral library search results are consistent with our parametric comparison tests in fifteen cases out of twenty-nine studied (52%).

## 7. Conclusion

This work updates, expands, and corrects the results of our long-term radar-driven observational campaign of main-belt asteroids (MBAs) focused on Bus-DeMeo Xc- and Xk-type objects (Tholen X and M class asteroids) using the Arecibo radar S-band telescope and NASA Infrared Telescope Facilities' SpeX instrument (Ockert-Bell et al., 2008; 2010; Shepard et al., 2008; 2010). Near simultaneous observations in the near-infrared and the S-band radar have revealed some interesting insights into the Tholen M-type (Bus-DeMeo Xc and Xk type) asteroids. By correlating near-infrared spectral features with radar derived metal abundances, we have cataloged the multi-wavelength behavior of 18 asteroids. We have also presented an analysis of the near-infrared properties of an additional 11 asteroids.

The near-infrared spectra of our asteroids exhibit a few weak silicate absorptions. These weak absorptions are not seen in spectra of pure metal meteorites (Cloutis et al., 2010). Such absorptions would not be expected if the asteroids were the parent bodies of pure metal meteorites. However, the steep near-infrared slope and moderate albedos of the asteroids suggest linkage with the pure metal meteorites. One meteorite was a match for many of the asteroids: the Landes iron meteorite. Landes has a large number of silicate inclusions in a nickel-iron matrix. The evidence indicates that the surfaces of these asteroids are composed of metal/silicate mixtures. We conclude that even asteroids with the highest radar metal abundances cannot be assumed to be pure metal.

Future work should complete a statistical sampling of these asteroids, strengthen laboratory databases on the properties of metal meteorites, and explore other wavelength regions.

**Acknowledgments**

J.R.N., B.E.C., and M.E.O.B. gratefully acknowledge support from NSF grants AST-0908098 and AST-0908217. E.A.C gratefully acknowledges support from NSERC, CFI, MRIF, and the University of Winnipeg. We thank the staff of the NASA Infrared Telescope Facility (IRTF), including Paul Sears, Bill Golisch, and Dave Griep for excellent telescope operation assistance. We thank Jonathan Joseph for programming assistance.

**Appendix A**

The following typographical errors were found in Ockert-Bell et al. (2010):
(1) The conversion from reflectance per micron to reflectance per $10^3$ angstroms for visual, NIR1, and NIR2 slopes was done incorrectly. Instead of dividing by 10, our slopes were multiplied by a factor of 10. We have resolved this issue, and reported slopes as reflectance per micron in this paper.
(2) Band 2 center and band 2 depth were switched for St. Mark's EC in Table 6b.
(3) The visual slope for Indarch EC was incorrect in Table 6b.
(4) A percent sign was included for 441 Bathilde in Table 3b.
(5) Incorrect NIR2 slope values were given for 21 Lutetia, 97 Klotho, 224 Oceana, 136 Austria, 441 Bathilde, and 758 Mancunia in Table 3a.

**Figures**

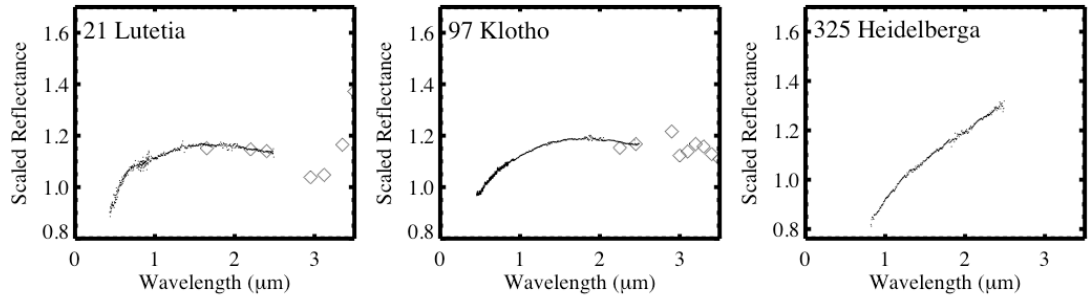

Fig. 1. Spectra of asteroids that do not have mafic absorptions at 0.9 and 1.9 μm . Visible (Chapman and Gaffey, 1979; Zellner et al., 1985; Bus and Binzel, 2002b; Fornasier et al., 2010) and thermal infrared data (Rivkin et al., (1995, 2000) and Rivkin (personal communication) for 97 Klotho) are added when possible. The Bus-DeMeo type, and the Rivkin type are given (W indicates that the asteroid has the 3-μm feature, M indicates that it does not).

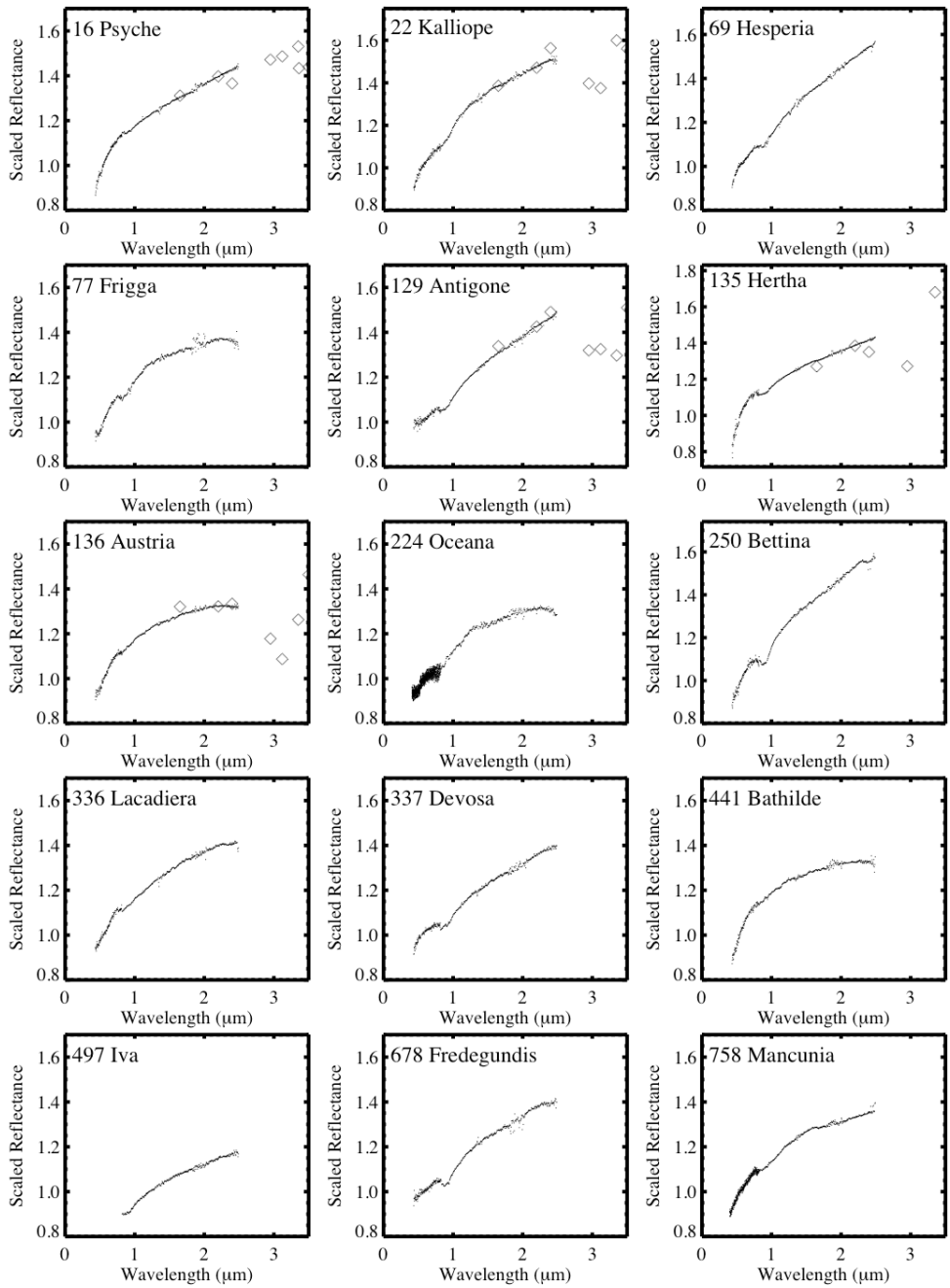

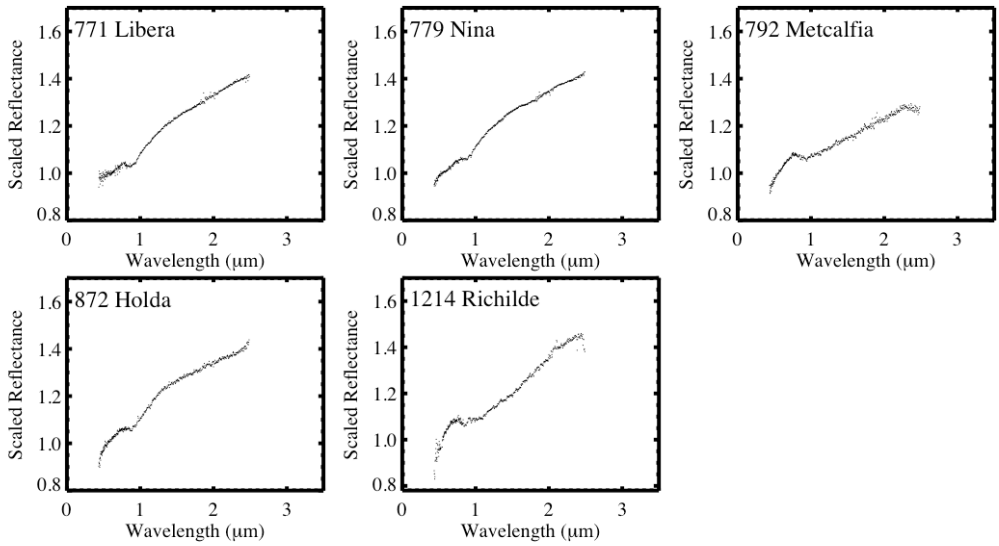

Fig. 2. Spectra of asteroids that have a measurable 0.9 μm mafic absorption but no 1.9 μm absorption.

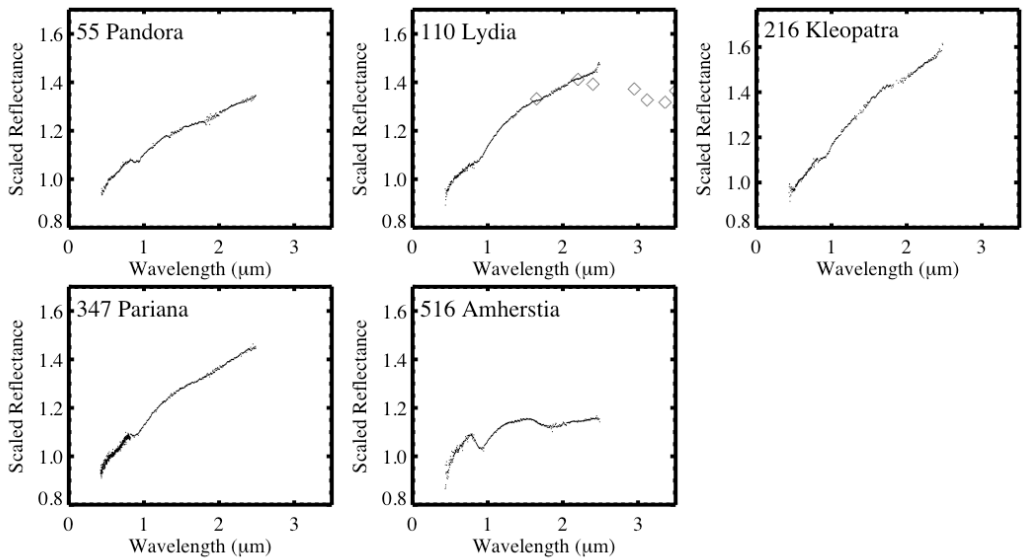

Fig. 3. Spectra of asteroids that have measurable 0.9 and 1.9 μm mafic absorptions.

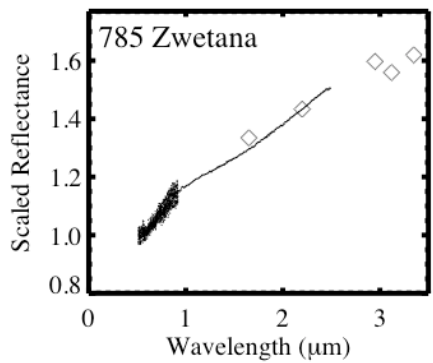

Fig. 4. A spectrum of 785 Zwetana.

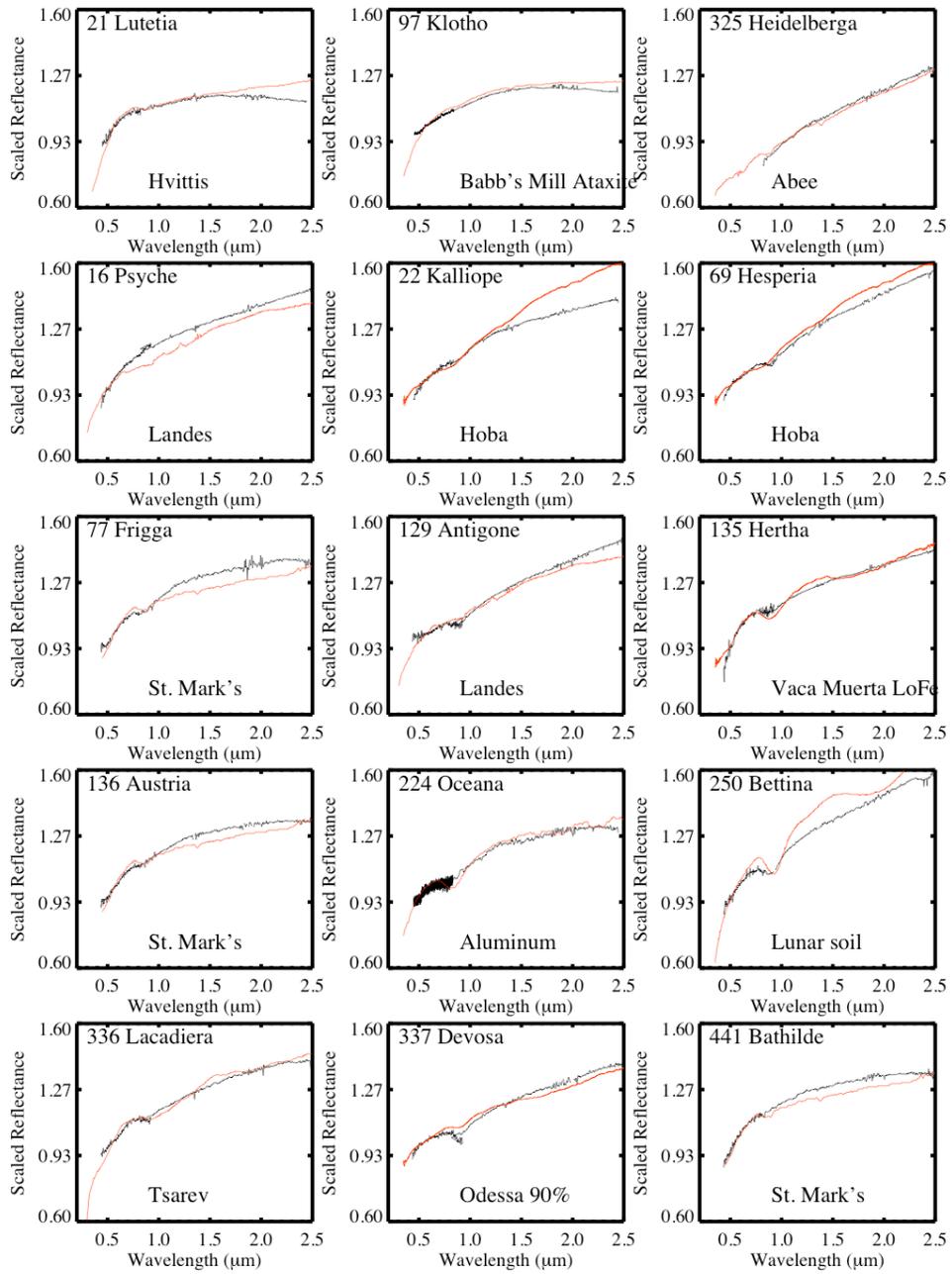

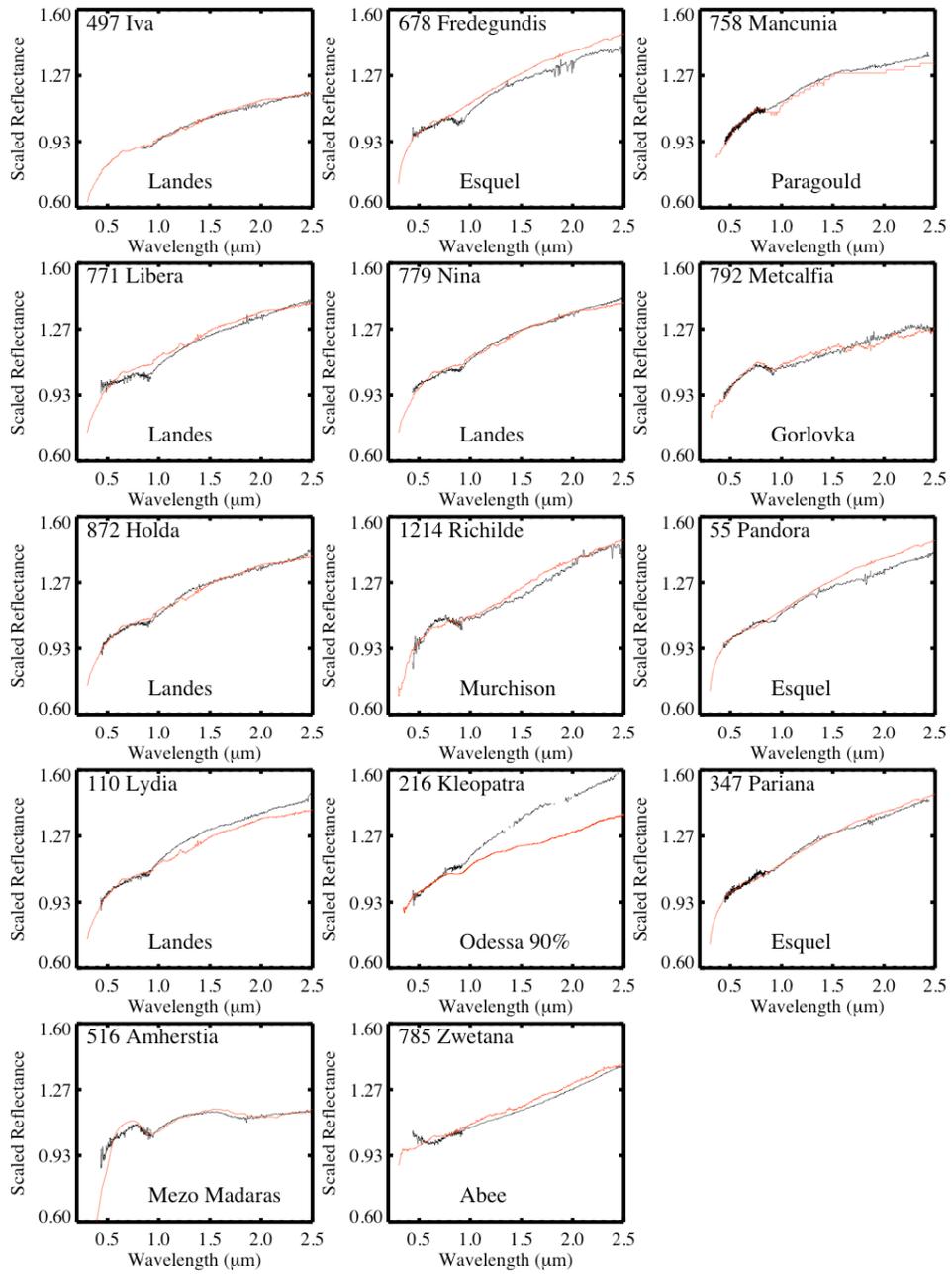

Fig. 5. The least-squares best-fit matches (red curve) found in the RELAB database for each asteroid (black curve)

**Tables**

**Table 1. Asteroids observed (2001-2011) for this study**

| Name | Bus-DeMeo type | Tholen type | $a$ (AU) | D (km) | P(hr) | Radar Albedo |
|---|---|---|---|---|---|---|
| 16 Psyche[+] | Xk | M | 2.92 | 186±30 | 4.196 | 0.42±0.10 |
| 21 Lutetia[+] | Xc | M | 2.44 | 100±11 | 8.172 | 0.24±0.07 |
| 22 Kalliope[+] | X | M | 2.91 | 162±3 | 4.148 | 0.18±0.05 |
| 55 Pandora[+] | Xk | M | 2.71 | 67 | 4.804 | -- |
| 69 Hesperia | Xk | M | 2.98 | 138±5 | 5.655 | 0.45±0.12 |
| 77 Frigga[+] | Xe | M | 2.67 | 69 | 9.012 | -- |
| 97 Klotho[+] | Xc | M | 2.67 | 83±5 | 35.15 | 0.26±0.05 |
| 110 Lydia[+] | Xk | M | 2.73 | 89±9 | 10.926 | 0.20±0.12 |
| 129 Antigone[+] | Xk | M | 2.87 | 113±12 | 4.957 | 0.36±0.09 |
| 135 Hertha[+] | Xk | M | 2.43 | 77±7 | 8.401 | 0.18±0.05 |
| 136 Austria[+] | Xc | M | 2.29 | 40 | 11.5 | -- |
| 216 Kleopatra[+] | Xe | M | 2.8 | 124±15 | 5.385 | 0.60±0.15 |
| 224 Oceana[+] | Xc | M | 2.65 | 62±2 | 9.388 | 0.25±0.10. |
| 250 Bettina[+] | Xk | M | 3.15 | 80 | 5.054 | -- |
| 325 Heidelberga | -- | M | 3.21 | 75±2 | 6.737 | 0.17±0.08 |
| 336 Lacadiera | Xk | D | 2.25 | 69±2 | 13.70 | -- |
| 337 Devosa | Xk | X | 2.38 | 59±2 | 4.65 | -- |
| 347 Pariana[+] | Xk | M | 2.61 | 51±5 | 4.053 | 0.36±0.09 |
| 441 Bathilde[+] | Xc | M | 2.81 | 70 | 10.447 | -- |
| 497 Iva[+] | Xk | M | 2.86 | 40±8 | 4.62 | 0.24±0.08 |
| 516 Amherstia | X | M | 2.68 | 73±2 | 7.49 | -- |
| 678 Fredegundis[+] | Xk | X | 2.57 | 42±4 | 11.62 | 0.18±0.05 |
| 758 Mancunia[+] | Xk | X | 3.19 | 85±7 | 12.738 | 0.55±0.14 |
| 771 Libera[+] | Xk | X | 2.65 | 29±2 | 5.892 | 0.17±0.04 |
| 779 Nina[+] | Xk | -- | 2.67 | 77±2 | 11.186 | 0.26±0.24 |
| 792 Metcalfia | X | -- | 2.62 | 61 | 9.17 | -- |
| 785 Zwetana[+] | Cb | M | 2.57 | 49±2 | 8.918 | 0.33±0.08 |
| 872 Holda[+] | Xk | M | 2.73 | 30 | 7.2 | -- |
| 1214 Richilde | Xk | -- | 2.71 | 35±3 | 9.860 | -- |

[+]Asteroids that were published in Ockert-Bell et al. (2008). Bus-DeMeo Taxonomy from DeMeo et al. (2009) or * retrieved from http://smass.mit.edu/busdemeoclass.html (March, 2009). *a* is the semi-major axis of the asteroid, D is the asteroid diameter, P is the rotation rate. Most values retrieved from the JPL Horizons database, http://ssd.jpl.nasa.gov (Jan, 2008). Diameter and radar albedo with error bars are from Shepard et al. (2010).

**Table 2. Asteroid Observational Circumstances**

| Name | UT Date | Time (UT) | Exposure Time (s) | Airmass | Standard Stars (airmass) |
|---|---|---|---|---|---|
| 69 Hesperia | 11 March 1 | 14:27 | - | 1.3 | C.(1.2) |
| 325 Heidelberga | 11 March 1 | 05:33, 05:59, 06:07 | 1080, 480, 960 | 1.1, 1.1, 1.1 | A.(1.0), D.(1.1) |
| 325 Heidelberga | 11 March 2 | 04:42, 05:22 | 720, 720 | 1.0 | A.(1.0), D.(1.1) |
| 325 Heidelberga | 11 March 3 | 05:10 | 1080 | 1.0 | A.(1.0), D.(1.1) |
| 336 Lacadiera | 11 August 25 | 09:40 | 180 | 1.1 | E.(1.1), G.(1.1) |
| 336 Lacadiera | 11 August 26 | 09:42 | 180 | 1.1 | E.(1.1), G.(1.1) |
| 336 Lacadiera | 11 August 27 | 09:25 | 270 | 1.1 | E.(1.1) |
| 337 Devosa | 11 August 25 | 14:10 | 1050 | 1.0 | D.(1.1) |
| 516 Amerstia | 11 March 1 | 07:40 | 360 | 1.0 | D.(1.1), B.(1.1) |
| 516 Amherstia | 11 March 2 | 07:15, 07:22, 07:55, 09:18 | 180, 180, 180, 180 | 1.0, 1.0, 1.0, 1.0 | D.(1.1), B.(1.1) |
| 792 Metcalfia | 11 August 25 | 07:48 | 720 | 1.2 | E.(1.1), F.(1.1) |
| 792 Metcalfia | 11 August 26 | 07:40, 07:53 | 240, 160 | 1.2, 1.2 | E.(1.1), F.(1.1) |
| 792 Metcalfia | 11 August 27 | 07:40 | 960 | 1.2 | E.(1.1), F.(1.1) |
| 1214 Richilde | 11 August 25 | 08:27 | 720 | 1.1 | E.(1.0), F.(1.1) |
| 1214 Richilde | 11 August 26 | 08:23 | 720 | 1.2 | E.(1.1), F.(1.1) |
| 1214 Richilde | 11 August 27 | 08:18 | 720 | 1.2 | E.(1.1) |

Standard stars-- A. 64Hyades, B. PG918+29C, C. SAO139464, Landolt stars: D. 97-249, E. 112-1333, F. 110-361, G. 113-276

**Table 3a. Asteroid Spectral Markers** (asteroids are organized according to presence of absorption bands)

| Name | NIR group | $P_v$ | VIS ±0.02 | NIR1 ± 0.01 | NIR2 ± 0.01 | Band Center 1 ±0.01 | Band Center 2 ±0.02 | NiFe Content |
|---|---|---|---|---|---|---|---|---|
| 21 Lutetia | 1 | 0.20 ±0.03 | 0.58 | 0.10 | -0.03 | - | - | Low |
| 97 Klotho | 1 | 0.23±0.03 | 0.37 | 0.12 | -0.02 | - | - | Low |
| 325 Heidelberga | 1 | 0.11 | - | 0.37 | 0.28 | - | - | Low |
| 16 Psyche | 2 | 0.23±0.05 | 0.67 | 0.19 | 0.17 | 0.91 | - | High |
| 22 Kalliope | 2 | 0.14±0.01 | 0.52 | 0.28 | 0.17 | 0.88 | - | Low |
| 69 Hesperia | 2 | 0.14 | 0.44 | 0.35 | 0.25 | 0.90 | - | High |
| 77 Frigga | 2 | 0.14 | 0.64 | 0.19 | 0.09 | 0.86 | - | - |
| 129 Antigone | 2 | 0.21±0.05 | 0.22 | 0.30 | 0.21 | 0.90 | - | High |
| 135 Hertha | 2 | 0.14±0.01 | 0.83 | 0.22 | 0.15 | 0.90 | - | Low |
| 136 Austria | 2 | 0.15 | 0.65 | 0.18 | 0.07 | 0.84* | - | - |
| 224 Oceana | 2 | 0.17±0.01 | 0.35 | 0.25 | 0.09 | 0.80 | - | Low |
| 250 Bettina | 2 | 0.26 | 0.63 | 0.35 | 0.27 | 0.89 | - | - |
| 336 Lacadiera | 2 | 0.05 | 0.58 | 0.14 | 0.15 | 0.89 | - | - |
| 337 Devosa | 2 | 0.16 | 0.30 | 0.19 | 0.11 | 0.92 | - | - |
| 441 Bathilde | 2 | 0.14 | 0.76 | 0.15 | 0.07 | 0.85* | - | - |
| 497 Iva | 2 | 0.13±0.03 | - | 0.24 | 0.14 | 0.91 | - | Low |
| 678 Fredegundis | 2 | 0.25±0.02 | 0.25 | 0.30 | 0.19 | 0.91 | - | Moderate |
| 758 Mancunia | 2 | 0.13±0.02 | 0.49 | 0.26 | 0.09 | 0.89 | - | High |
| 771 Libera | 2 | 0.13±0.01 | 0.19 | 0.30 | 0.19 | 0.90 | - | Low |
| 779 Nina | 2 | 0.14±0.01 | 0.28 | 0.28 | 0.17 | 0.91 | - | High |
| 792 Metcalfia | 2 | 0.04 | 0.45 | 0.18 | 0.17 | 0.96 | - | - |
| 872 Holda | 2 | 0.21 | 0.38 | 0.32 | 0.19 | 0.92 | - | - |
| 1214 Richilde | 2 | 0.06 | 0.57 | 0.30 | 0.35 | 0.95 | - | - |
| 55 Pandora | 3 | 0.30 | 0.37 | 0.19 | 0.15 | 0.93 | 1.87 | - |
| 110 Lydia | 3 | 0.16±0.02 | 0.34 | 0.30 | 0.17 | 0.89 | 1.72 | Moderate |
| 216 Kleopatra | 3 | 0.12±0.02 | 0.40 | 0.36 | 0.20 | 0.92 | 1.95 | High |
| 347 Pariana | 3 | 0.18±0.02 | 0.35 | 0.28 | 0.19 | 0.90 | 1.83 | High |
| 516 Amherstia | 3 | 0.16 | 0.59 | 0.13 | 0.02 | 0.93 | 1.85 | - |
| 785 Zwetana | 4 | 0.12±0.01 | 0.37 | 0.18 | 0.25 | -- | -- | High |

$P_v$ is the geometric albedo from the JPL Horizons database or Shepard et al. (2010). Visible spectrum (VIS) slope is measured from 0.45 to 0.7 μm, the near-infrared slopes are measured between: NIR1 1.1 - 1.45 μm, NIR2 1.6 - 2.3 μm. Slope units are Reflectance/μm. The NiFe content and Radar Analog are from Shepard et al. (2010).
* The band centers for these two asteroids do not correspond to typical minerals found on asteroid surfaces.

**Table 3b. Asteroid Band Depth Determination**

| Name | Band 1 depth (±0.003 μm) | Band 1 SSR | Band 2 depth (±0.002 μm) | Band 2 SSR |
|---|---|---|---|---|
| 21 Lutetia | 0 | 0 | 0 | 0 |
| 97 Klotho | 0 | 0 | 0 | 0 |
| 325 Heidelberga | 0 | 0 | 0 | 0 |
| | | | | |
| 16 Psyche | 0.013 | 6.5 | 0 | 0 |
| 22 Kalliope | 0.029 | 5.8 | 0 | 0 |
| 69 Hesperia | 0.033 | 8.3 | 0 | 0 |
| 77 Frigga | 0.031 | 6.9 | 0 | 0 |
| 129 Antigone | 0.038 | 9.6 | 0 | 0 |
| 135 Hertha | 0.030 | 7.3 | 0 | 0 |
| 136 Austria | 0.012 | 3.5 | 0 | 0 |
| 224 Oceana | 0.052 | 3.9 | 0 | 0 |
| 250 Bettina | 0.062 | 8.9 | 0 | 0 |
| 336 Lacadiera | 0.030 | 5.1 | 0 | 0 |
| 337 Devosa | 0.044 | 7.2 | 0 | 0 |
| 441 Bathilde | 0.015 | 4.1 | 0 | 0 |
| 497 Iva | 0.020 | 6.2 | 0 | 0 |
| 678 Fredegundis | 0.054 | 11.9 | 0 | 0 |
| 758 Mancunia | 0.029 | 5.6 | 0 | 0 |
| 771 Libera | 0.044 | 12.5 | 0 | 0 |
| 779 Nina | 0.030 | 8.2 | 0 | 0 |
| 792 Metcalfia | 0.030 | 6.4 | 0 | 0 |
| 872 Holda | 0.038 | 7.9 | 0 | 0 |
| 1214 Richilde | 0.044 | 4.9 | 0 | 0 |
| | | | | |
| 55 Pandora | 0.025 | 7.3 | 0.014 | 3.4 |
| 110 Lydia | 0.034 | 9.1 | 0.008 | 3.6 |
| 216 Kleopatra | 0.028 | 8.4 | 0.019 | 6.1 |
| 347 Pariana | 0.030 | 6.7 | 0.008 | 3.4 |
| 516 Amherstia | 0.060 | 8.1 | 0.029 | 9.8 |
| | | | | |
| 785 Zwetana | 0 | 0 | 0 | 0 |

See text for explanation of Band Signal Strength Ratio (SSR).

**Table 4. Best-fit Meteorite Type: RELAB Analog Method**

| Name | $P_v$ | Meteorite analog (norm. at 0.55 µm) | Sample RELAB ID | Sample Reflectance at 0.55 µm | $\chi^2$ |
|---|---|---|---|---|---|
| 21 Lutetia | 0.20 | Hvittis EC | MR-MJG-024 | 0.18 | 0.01 |
| 97 Klotho | 0.23 | Babb's Mill Ataxite | MR-MJG-083 | 0.24 | 0.001 |
| 325 Heidelberga | 0.11 | Abee EC | MR-MJG-020 | 0.10 | 0.0003 |
| | | | | | |
| 16 Psyche | 0.23 | Landes IM | MB-TXH-046 | 0.17 | 0.002 |
| 22 Kalliope | 0.14 | Hoba Ataxite | EAC | 0.10 | 0.02 |
| 69 Hesperia | 0.14 | Hoba Ataxite | EAC | 0.10 | 0.001 |
| 77 Frigga | 0.14 | St. Mark's EC E5 | MR-MJG-022 | 0.08 | 0.009 |
| 129 Antigone | 0.21 | Landes IM | MB-TXH-046 | 0.17 | 0.001 |
| 135 Hertha | 0.14 | Vaca Muerta | EAC | 0.10 | 0.0006 |
| 136 Austria | 0.15 | St. Mark's EC E5 | MR-MJG-022 | 0.09 | 0.0004 |
| 224 Oceana | 0.17 | Aluminum | SI-PHS-001 | 0.19 | 0.01 |
| 250 Bettina | 0.26 | Lunar soil | LS-JBA-155-N2 | 0.25 | 0.02 |
| 336 Lacadiera | 0.05 | Tsarev OC | MA-ATB-055 | 0.05 | 0.0005 |
| 337 Devosa | 0.16 | Odessa 90% IM | EAC | 0.10 | 0.02 |
| 441 Bathilde | 0.14 | St. Mark's EC E5 | MR-MJG-022 | 0.09 | 0.0005 |
| 497 Iva | 0.13 | Landes IM | MB-TXH-046 | 0.17 | 0.0001 |
| 678 Fredegundis | 0.25 | Esquel SIM | MB-TXH-043 | 0.30 | 0.001 |
| 758 Mancunia | 0.13 | Paragould OC | MR-MJG-076 | 0.06 | 0.0003 |
| 771 Libera | 0.13 | Landes IM | MB-TXH-046 | 0.17 | 0.0009 |
| 779 Nina | 0.14 | Landes IM | MB-TXH-046 | 0.17 | 0.0002 |
| 792 Metcalfia | 0.03 | Gorlovka OC | RS-CMP-048 | 0.10 | 0.004 |
| 872 Holda | 0.21 | Landes IM | MB-TXH-046 | 0.17 | 0.0002 |
| 1214 Richilde | 0.06 | Murchison CM | MB-TXH-064 | 0.03 | 0.0008 |
| | | | | | |
| 55 Pandora | 0.30 | Esquel SIM | MB-TXH-043 | 0.30 | 0.0009 |
| 110 Lydia | 0.16 | Landes IM | MB-TXH-046 | 0.17 | 0.001 |
| 216 Kleopatra | 0.12 | Odessa 90% IM | EAC | 0.10 | 0.1 |
| 347 Pariana | 0.18 | Esquel SIM | MB-TXH-043-H | 0.14 | 0.0002 |
| 516 Amherstia | 0.16 | Meso Maderas OC | TB-TJM-079 | 0.16 | 0.001 |
| | | | | | |
| 785 Zwetana | 0.12 | Abee EC | MB-TXH-040-F | 0.07 | 0.0006 |

**Table 5: Meteorite Parameter Ranges**

| Meteorite type | REF range | VIS range | NIR1 range | NIR2 range | Band 1 Center range | Band 2 Center range |
|---|---|---|---|---|---|---|
| CI | 0.02-0.11 | -0.78 – 1.98 | -0.20 – 0.60 | -0.09 – 0.33 | 0.71 – 1.29 | 1.84 – 2.19 |
| CK | 0.09 – 0.22 | -0.12 – 0.55 | -0.09 – 0.55 | -0.12 – 0.02 | 0.96 – 1.13 | 1.95 – 2.07* |
| CM | 0.03 – 0.09 | -0.09 – 1.68 | -0.08 – 1.23 | -0.11 – 0.60 | 0.49 – 1.21 | 1.39 – 2.53 |
| CO | 0.05 – 0.19 | 0.50 – 1.86 | -0.11 – 0.31 | -0.17 – 0.17 | 0.93 – 1.22 | 1.91 – 2.25 |
| CR | 0.04 – 0.13 | 0.61 – 2.65 | 0.0 – 0.29 | -0.04 – 0.14 | 0.92 – 1.27* | 1.89 – 2.30* |
| CV | 0.05 – 0.14 | -0.02 – 2.58 | -0.11 – 0.35 | -0.23 – 0.14 | 0.86 – 1.16 | 1.90 – 2.14* |
| EC | 0.06 – 0.18 | 0.23 – 2.05 | 0.08 – 0.25 | 0.06 – 0.27 | 0.81 – 0.90 | 1.87 – 2.26* |
| IM | 0.04 – 0.41 | 0.09 – 2.08 | 0.10 – 0.49 | 0.02 – 0.34 | 0.65 – 0.96* | 1.96 – 2.25* |
| SIM | 0.09 – 0.17 | 0.16 – 1.19 | -0.03 – 0.30 | -0.12 – 0.31 | 0.77 – 0.91* | 1.76 – 2.01* |
| UR | 0.10 – 0.18 | 1.18 – 3.89 | -0.11 – 0.26 | -0.07 – 0.06 | 0.92 – 1.30 | 1.89 – 2.09* |

REF is the reflectance at 0.55 μm
* indicates that some meteorites in the group do not show presence of a band

**Table 6. Best-fit Meteorite Type: Parameter Range Comparison Test**

| Name | # of Parameters that fit (out of 6) | | | | | | | | | | Best-fit type |
|---|---|---|---|---|---|---|---|---|---|---|---|
| | CI | CK | CM | CO | CR | CV | EC | IM | SIM | UR | |
| 21 Lutetia | 3 | 3 | 2 | 2 | 3 | 3 | 2 | 4 | 4 | 2 | IM/SIM |
| 97 Klotho | 3 | 4 | 3 | 2 | 4 | 4 | 3 | 5 | 5 | 3 | IM/SIM |
| 325 Heidelberga | 4 | 4 | 3 | 1 | 3 | 3 | 2 | 5 | 4 | 2 | IM |
| | | | | | | | | | | | |
| 16 Psyche | 4 | 2 | 4 | 2 | 4 | 4 | 4 | 6 | 4 | 2 | IM |
| 22 Kalliope | 4 | 4 | 4 | 3 | 3 | 4 | 5 | 6 | 6 | 2 | IM/SIM |
| 69 Hesperia | 3 | 3 | 3 | 1 | 2 | 3 | 4 | 5 | 4 | 2 | IM |
| 77 Frigga | 4 | 3 | 4 | 4 | 5 | 5 | 6 | 6 | 6 | 3 | EC/IM/SIM |
| 129 Antigone | 4 | 4 | 4 | 1 | 2 | 4 | 3 | 6 | 5 | 1 | IM |
| 135 Hertha | 4 | 3 | 4 | 4 | 4 | 4 | 6 | 6 | 6 | 3 | EC/IM/SIM |
| 136 Austria | 4 | 3 | 4 | 4 | 5 | 4 | 6 | 6 | 6 | 3 | EC/IM/SIM |
| 224 Oceana | 4 | 4 | 4 | 3 | 4 | 4 | 4 | 6 | 6 | 3 | IM/SIM |
| 250 Bettina | 4 | 2 | 4 | 1 | 3 | 4 | 4 | 6 | 4 | 1 | IM |
| 336 Lacadiera | 5 | 2 | 5 | 3 | 5 | 4 | 5 | 6 | 5 | 2 | IM |
| 337 Devosa | 4 | 4 | 4 | 3 | 4 | 5 | 5 | 6 | 6 | 3 | IM/SIM |
| 441 Bathilde | 4 | 3 | 4 | 4 | 5 | 4 | 6 | 6 | 6 | 3 | EC/IM/SIM |
| 497 Iva | 4 | 4 | 4 | 3 | 5 | 6 | 4 | 5 | 4 | 3 | CV |
| 678 Fredegundis | 4 | 3 | 4 | 1 | 2 | 4 | 3 | 6 | 4 | 1 | IM |
| 758 Mancunia | 4 | 4 | 4 | 3 | 5 | 6 | 5 | 6 | 6 | 3 | CV/IM/SIM |
| 771 Libera | 4 | 4 | 4 | 2 | 3 | 5 | 4 | 6 | 5 | 2 | IM |
| 779 Nina | 4 | 4 | 4 | 2 | 3 | 4 | 4 | 6 | 6 | 2 | IM/SIM |
| 792 Metcalfia | 5 | 4 | 5 | 3 | 3 | 4 | 4 | 4 | 4 | 3 | CI/CM |
| 872 Holda | 4 | 4 | 4 | 0 | 2 | 4 | 3 | 6 | 3 | 1 | IM |
| 1214 Richilde | 5 | 2 | 5 | 4 | 4 | 5 | 4 | 6 | 3 | 3 | IM |
| | | | | | | | | | | | |
| 55 Pandora | 4 | 3 | 5 | 3 | 3 | 4 | 4 | 6 | 4 | 3 | IM |
| 110 Lydia | 4 | 4 | 5 | 3 | 2 | 4 | 5 | 6 | 5 | 2 | IM |
| 216 Kleopatra | 4 | 4 | 5 | 2 | 3 | 4 | 4 | 6 | 4 | 2 | IM |
| 347 Pariana | 4 | 4 | 5 | 2 | 3 | 4 | 4 | 6 | 5 | 2 | IM |
| 516 Amherstia | 4 | 4 | 5 | 4 | 4 | 5 | 4 | 6 | 5 | 5 | IM |
| | | | | | | | | | | | |
| 785 Zwetana | 3 | 4 | 5 | 2 | 4 | 4 | 5 | 6 | 6 | 3 | IM/SIM |

EC = enstatite chondrite, IM = iron meteorite, SIM = stony-iron meteorite, CI, CK, CM, CO, CR, CV = carbonaceous chondrite types

**Table 7. Best-fit Meteorite Type: Average, Mean, and Median Distance Tests**

| Names | Average Best-fit type | Mean Best-fit type | Median Best-fit type |
|---|---|---|---|
| 21 Lutetia | IM | IM | IM |
| 97 Klotho | IM | IM | CK |
| 325 Heidelberga | IM | IM | IM |
| | | | |
| 16 Psyche | IM | IM | EC |
| 22 Kalliope | CM/IM | IM | IM |
| 69 Hesperia | IM | SIM | IM |
| 77 Frigga | EC/IM | SIM | EC |
| 129 Antigone | IM | IM | IM |
| 135 Hertha | CR | IM | EC |
| 136 Austria | IM | SIM | EC |
| 224 Oceana | CM/IM | IM | IM |
| 250 Bettina | IM | IM | IM |
| 336 Lacadiera | CR/SIM | IM | IM |
| 337 Devosa | CR/IM | IM | CK |
| 441 Bathilde | CK/SIM | SIM | EC |
| 497 Iva | CR/SIM | SIM | CK |
| 678 Fredegundis | IM | IM | IM |
| 758 Mancunia | CM/CR | SIM | IM |
| 771 Libera | IM | IM | IM |
| 779 Nina | CK/CM | IM | CK |
| 792 Metcalfia | CM | IM | IM |
| 872 Holda | IM | IM | IM |
| 1214 Richilde | CR/IM | IM | IM |
| | | | |
| 55 Pandora | CR | SIM | SIM |
| 110 Lydia | CM | SIM | SIM |
| 216 Kleopatra | CI/IM | SIM | CM |
| 347 Pariana | IM | SIM | SIM |
| 516 Amherstia | - | SIM | SIM |
| | | | |
| 785 Zwetana | IM | SIM | SIM |

**Table 8a. Summary of Meteorite Classification Using Discriminant Analysis**

|  | **True Group** | | | | | | |
|---|---|---|---|---|---|---|---|
| **Predicted Group** | CICM | CKCOCV | CR | EC | IM | SIM | UR |
| CICM | 121 | 6 | 0 | 0 | 0 | 0 | 0 |
| CKCOCV | 7 | 72 | 1 | 0 | 0 | 0 | 0 |
| CR | 2 | 1 | 8 | 1 | 2 | 0 | 1 |
| EC | 1 | 6 | 1 | 6 | 2 | 0 | 0 |
| IM | 0 | 0 | 1 | 0 | 17 | 1 | 0 |
| SIM | 1 | 0 | 0 | 2 | 2 | 4 | 0 |
| UR | 0 | 0 | 4 | 0 | 0 | 0 | 18 |
| Total Number | 132 | 85 | 15 | 9 | 23 | 5 | 19 |
| Number Correct | 121 | 72 | 8 | 6 | 17 | 4 | 18 |
| Percentage | 91.7% | 84.5% | 53.3% | 66.7% | 73.9% | 80.0% | 94.7% |
| **Overall correct** | **85.4%** | | | | | | |

**Table 8b. Best-fit Meteorite Type: Discriminant Analysis Method**

| Name | Best-fit type | Probability (%) |
|---|---|---|
| 21 Lutetia | IM | 94 |
| 97 Klotho | IM | 99 |
| 325 Heidelberga | IM | 98 |
| | | |
| 16 Psyche | IM | 88 |
| 22 Kalliope | IM | 53 |
| 69 Hesperia | SIM | 58 |
| 77 Frigga | EC | 64 |
| 129 Antigone | IM | 93 |
| 135 Hertha | EC | 70 |
| 136 Austria | EC | 65 |
| 224 Oceana | IM | 67 |
| 250 Bettina | IM | 98 |
| 336 Lacadiera | EC | 73 |
| 337 Devosa | IM | 48 |
| 441 Bathilde | EC | 77 |
| 497 Iva | EC | 49 |
| 678 Fredegundis | IM | 97 |
| 758 Mancunia | EC | 70 |
| 771 Libera | IM | 49 |
| 779 Nina | IM | 48 |
| 792 Metcalfia | EC | 75 |
| 872 Holda | IM | 90 |
| 1214 Richilde | EC | 85 |
| | | |
| 55 Pandora | SIM | 79 |
| 110 Lydia | SIM | 81 |
| 216 Kleopatra | SIM | 41 |
| 347 Pariana | SIM | 89 |
| 516 Amherstia | SIM | 48 |
| | | |
| 785 Zwetana | SIM | 93 |

For probabilities below 50% we conclude that this method is not reliable.

**Table 9. Best-fit Meteorite Type: All Methods of Parametric Comparison**

| Name | Range Method | Average Method | Mean Method | Median Method | Discriminant Analysis | Summary of PC Methods | RELAB Analog |
|---|---|---|---|---|---|---|---|
| 21 Lutetia | IM/SIM | IM | IM | IM | IM | IM | EC |
| 97 Klotho | IM/SIM | IM | IM | CK | IM | IM | IM |
| 325 Heidelberga | IM | IM | IM | IM | IM | IM | EC |
|  |  |  |  |  |  |  |  |
| 16 Psyche | IM | IM | IM | EC | IM | IM | IM |
| 22 Kalliope | IM/SIM | CM/IM | IM | IM | IM/EC | IM | IM |
| 69 Hesperia | IM | IM | SIM | IM | SIM | IM | IM |
| 77 Frigga | EC/IM/SIM | EC/IM | SIM | EC | EC | EC | EC |
| 129 Antigone | IM | IM | IM | IM | IM | IM | IM |
| 135 Hertha | EC/IM/SIM | CR | IM | EC | EC | EC | EC |
| 136 Austria | EC/IM/SIM | IM | SIM | EC | EC | EC | EC |
| 224 Oceana | IM/SIM | CM/IM | IM | IM | IM | IM | Aluminum |
| 250 Bettina | IM | IM | IM | IM | IM | IM | Lunar soil |
| 336 Lacadiera | IM | CR/SIM | IM | IM | EC | IM | OC |
| 337 Devosa | IM/SIM | CR/IM | IM | CK | IM/EC | IM | IM |
| 441 Bathilde | EC/IM/SIM | CK/SIM | SIM | EC | EC | EC | EC |
| 497 Iva | CV | CR/SIM | SIM | CK | EC/IM | - | IM |
| 678 Fredegundis | IM | IM | IM | IM | IM | IM | EC |
| 758 Mancunia | CV/IM/SIM | CM/CR | SIM | IM | EC | - | OC |
| 771 Libera | IM | IM | IM | IM | IM/EC | IM | IM |
| 779 Nina | IM/SIM | CK/CM | IM | CK | IM/EC | IM | IM |
| 792 Metcalfia | CI/CM | CM | IM | IM | EC | - | OC |
| 872 Holda | IM | IM | IM | IM | IM | IM | IM |
| 1214 Richilde | IM | CR/IM | IM | IM | EC | IM | CM |
|  |  |  |  |  |  |  |  |
| 55 Pandora | IM | CR | SIM | SIM | SIM | SIM | SIM |
| 110 Lydia | IM | CM | SIM | SIM | SIM | SIM | IM |
| 216 Kleopatra | IM | CI/IM | SIM | CM | SIM/CICM | - | IM |
| 347 Pariana | IM | IM | SIM | SIM | SIM | SIM | SIM |
| 516 Amherstia | IM | - | SIM | SIM | SIM/CKCOCV | SIM | OC |
|  |  |  |  |  |  |  |  |
| 785 Zwetana | IM/SIM | IM | SIM | SIM | SIM | SIM | EC |

**Table 10. Best-fit Meteorite Type for All Methods Compared to Radar Analog**

| Name | RELAB Analog | Summary of PC Methods | Radar Analog | NiFe Content |
|---|---|---|---|---|
| **21 Lutetia** | **EC** | **IM** | **EC,CH** | **Low** |
| 97 Klotho | IM | IM | EC,CH | Low |
| **325 Heidelberga** | **EC** | **IM** | **SIM,EC** | **Low** |
| | | | | |
| **16 Psyche** | **IM** | **IM** | **IM,CB** | **High** |
| 22 Kalliope | IM | IM | CH,EC | Low |
| **69 Hesperia** | **IM** | **IM** | **IM** | **High** |
| 129 Antigone | IM | IM | CB? | High |
| **135 Hertha** | **EC** | **EC** | **CH,EC** | **Low** |
| 224 Oceana | Aluminum | IM | EC,CH | Low |
| 497 Iva | IM | - | SIM,CH | Low |
| 678 Fredegundis | EC | IM | SIM,CB | Moderate |
| 758 Mancunia | OC | - | IM,CB | High |
| 771 Libera | IM | IM | SIM,CH | Low |
| **779 Nina** | **IM** | **IM** | **IM,CB** | **High** |
| | | | | |
| 110 Lydia | IM | SIM | CB,CH | Moderate |
| **216 Kleopatra** | **IM** | **-** | **IM** | **High** |
| 347 Pariana | SIM | SIM | IM,CB | High |
| | | | | |
| 785 Zwetana | EC | SIM | IM,CB | High |

**Bold** font indicates objects for which agreement was found between the radar analog type and the best-fit type for PC and/or RELAB method.